\documentclass[twocolumn]{aastex63}
\usepackage{amsmath}

\newcommand{\msun}{${\rm M_{\sun}}$}

\def\ltsima{$\; \buildrel < \over \sim \;$}
\def\simlt{\lower.5ex\hbox{\ltsima}}
\def\gtsima{$\; \buildrel > \over \sim \;$}
\def\simgt{\lower.5ex\hbox{\gtsima}}
\def\kms{{\rm\,km\,s^{-1}}}

\def\kpc{{\rm\,kpc}}

\def\msun{{\rm\,M_\odot}}

\def\g{{\rm\,g}}

\def \r {r$^{1/4}$ }


\def\s{\ifmmode \widetilde \else \~\fi}
\def\={\overline}

\def\spose#1{\hbox to 0pt{#1\hss}}

\def\lta{\mathrel{\spose{\lower 3pt\hbox{$\mathchar"218$}}
     \raise 2.0pt\hbox{$\mathchar"13C$}}}
\def\gta{\mathrel{\spose{\lower 3pt\hbox{$\mathchar"218$}}
     \raise 2.0pt\hbox{$\mathchar"13E$}}}
\def\Dt{\spose{\raise 1.5ex\hbox{\hskip3pt$\mathchar"201$}}}    
\def\dt{\spose{\raise 1.0ex\hbox{\hskip2pt$\mathchar"201$}}}    

\def\r{${\rm r^{1/4}}$}

\def\dotsfill{\leaders\hbox to 1em{\hss.\hss}\hfill}
\def\sun{\odot}
\def\earth{\oplus}

\def\Gyr{{\rm\,Gyr}}

\def\ltsima{$\; \buildrel < \over \sim \;$}
\def\gtsima{$\; \buildrel > \over \sim \;$}
\def\lsim{\lower.5ex\hbox{\ltsima}}
\def\gsim{\lower.5ex\hbox{\gtsima}}
\def\lapp{\ifmmode\stackrel{<}{_{\sim}}\else$\stackrel{<}{_{\sim}}$\fi}
\def\gapp{\ifmmode\stackrel{>}{_{\sim}}\else$\stackrel{<}{_{\sim}}$\fi}

\defcitealias{V21}{V21}
\usepackage{subfigure}

\submitjournal{ApJ}

\shorttitle{Antaeus: a retrograde group of tidal debris in the MW disk plane}
\shortauthors{Oria et al.}
\graphicspath{{./}{figures/}}

\begin{document}

\title{Antaeus: a retrograde group of tidal debris in the Milky Way's disk plane}

\correspondingauthor{Pierre-Antoine Oria}
\email{pierre-antoine.oria@astro.unistra.fr}

\author{Pierre-Antoine Oria}
\affiliation{Universit\'e de Strasbourg, CNRS, Observatoire astronomique de Strasbourg, UMR 7550, F-67000 Strasbourg, France}

\author{Wassim Tenachi}
\affiliation{Universit\'e de Strasbourg, CNRS, Observatoire astronomique de Strasbourg, UMR 7550, F-67000 Strasbourg, France}

\author[0000-0002-3292-9709]{Rodrigo Ibata}
\affiliation{Universit\'e de Strasbourg, CNRS, Observatoire astronomique de Strasbourg, UMR 7550, F-67000 Strasbourg, France}

\author[0000-0003-3180-9825]{Benoit Famaey}
\affiliation{Universit\'e de Strasbourg, CNRS, Observatoire astronomique de Strasbourg, UMR 7550, F-67000 Strasbourg, France}

\author{Zhen Yuan}
\affiliation{Universit\'e de Strasbourg, CNRS, Observatoire astronomique de Strasbourg, UMR 7550, F-67000 Strasbourg, France}

\author{Anke Arentsen}
\affiliation{Universit\'e de Strasbourg, CNRS, Observatoire astronomique de Strasbourg, UMR 7550, F-67000 Strasbourg, France}

\author{Nicolas Martin}
\affiliation{Universit\'e de Strasbourg, CNRS, Observatoire astronomique de Strasbourg, UMR 7550, F-67000 Strasbourg, France}

\author{Akshara Viswanathan}
\affiliation{Kapteyn Astronomical Institute, University of Groningen, Landleven 12, NL-9747AD Groningen, the Netherlands}

\begin{abstract}
We present the discovery of a wide retrograde moving group in the disk plane of the Milky Way using action-angle coordinates derived from the \textit{Gaia} DR3 catalog. The structure is identified from a sample of its members that are currently almost at the pericenter of their orbit and are passing through the Solar neighborhood. The motions of the stars in this group are highly correlated, indicating that the system is probably not phase mixed. With a width of at least $1.5\kpc$ and with a probable intrinsic spread in metallicity, this structure is most likely the wide remnant of a tidal stream of a disrupted ancient dwarf galaxy (age $\sim 12\Gyr$, $\langle {\rm [Fe/H]} \rangle \sim -1.74$). The structure presents many similarities (e.g. in energy, angular momentum, metallicity, and eccentricity) with the Sequoia merging event. However, it possesses extremely low vertical action $J_z$ which makes it unique even amongst Sequoia dynamical groups. As the low $J_z$ may be attributable to dynamical friction, we speculate that the these stars may be the remnants of the dense core of the Sequoia progenitor.
\end{abstract}

\keywords{Galactic Archeology --- galaxies: disk --- galaxies: kinematics and dynamics --- Local Group}

\section{Introduction}
\label{sec:Introduction}

The complex formation and merging history of the Milky Way (MW) can perhaps be best understood by examining its stellar halo, host to many tidal debris of disrupted galaxies and globular clusters. Dynamical times in the halo are long, so the debris can persist there as coherent phase space structures for billions of years (see e.g. \citealt{2000MNRAS.319..657H}), making them easier for us to detect.

With the advent of the \textit{Gaia} mission \citep{2016A&A...595A...1G} and its superb astrometric data, the task of digging into the stellar halo to uncover the past has been made more accessible. The stellar halo of the MW is now understood to be the product of several important accretion events making up most of its population \citep{2019A&A...632A...4D}, the biggest of which being Gaia-Sausage/Enceladus \citep{2018MNRAS.478..611B,2018Natur.563...85H}. Stream finding algorithms \citep{2018MNRAS.481.3442M,2021ApJ...914..123I} have now detected dozens of kinematically coherent structures which will help chart the acceleration field of our Galaxy, providing a wealth of model-agnostic information. 

The \textit{Gaia} data also makes it possible to use action coordinates $(J_r,J_\phi,J_z)$ to detect stellar structures. Actions keep relevance over very long times if the potential evolves slowly and are thus especially useful to trace past mergers. Recently, \citet{2020ApJ...891...39Y}, \citet{2020ApJ...901...48N} and \citet{2022ApJ...926..107M}
used these quantities to detect and construct maps of the MW's dynamical groups and link them to important merger events.  

\begin{figure*}[!htb]
\centering
  \includegraphics[angle=0,  clip, width=\textwidth]{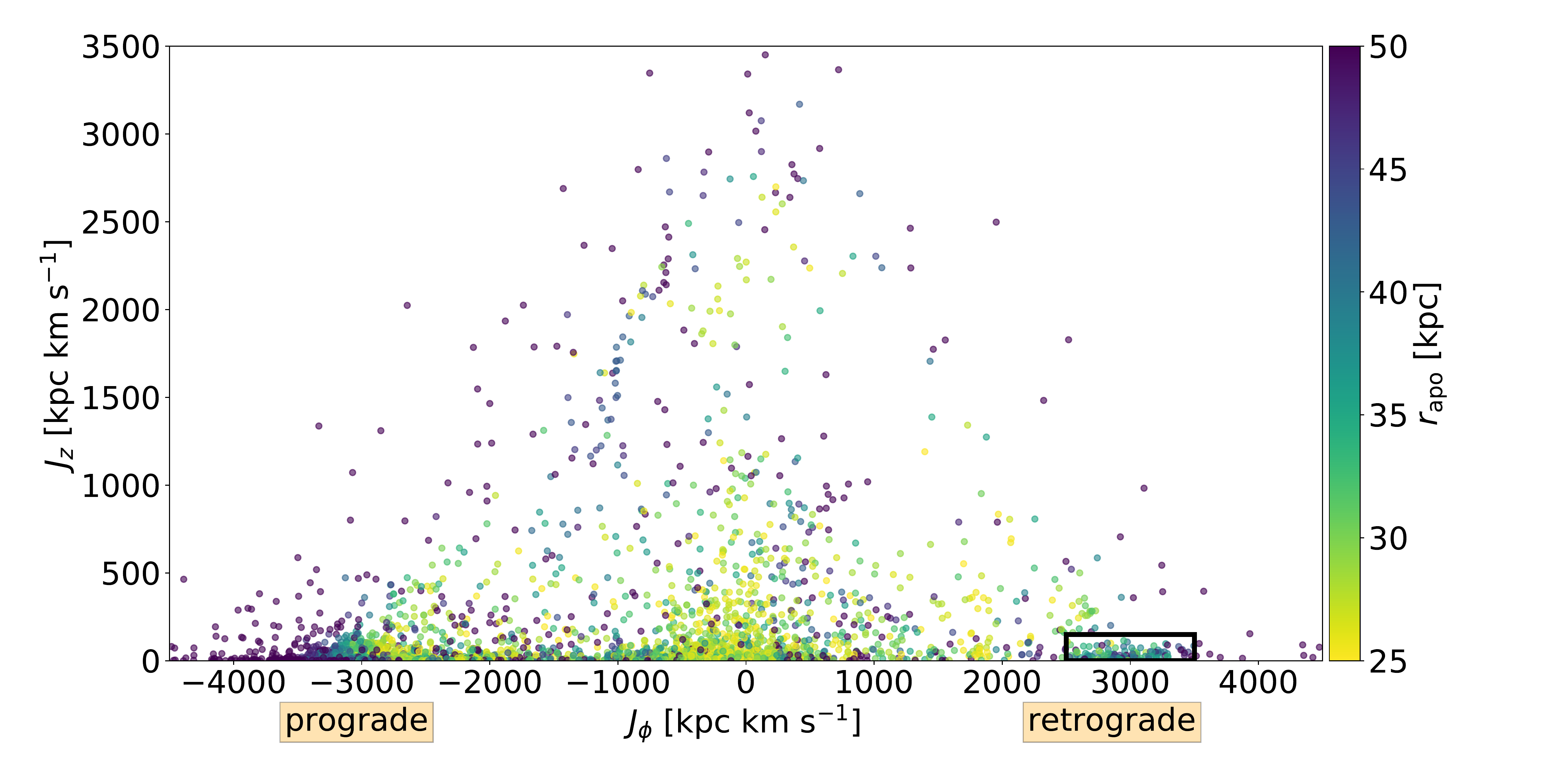}
  \includegraphics[angle=0,  clip, width=\textwidth]{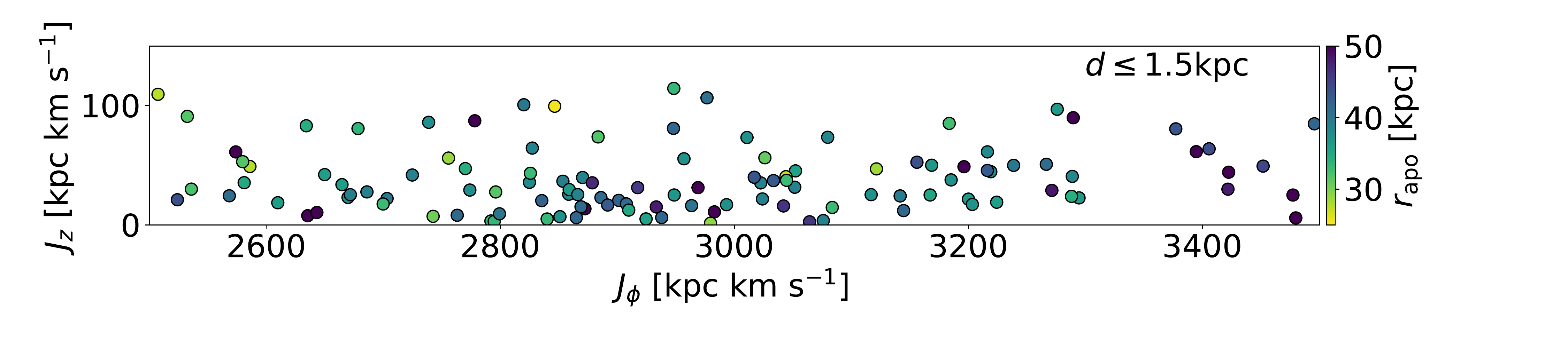}
  \includegraphics[angle=0,  clip, width=\textwidth]{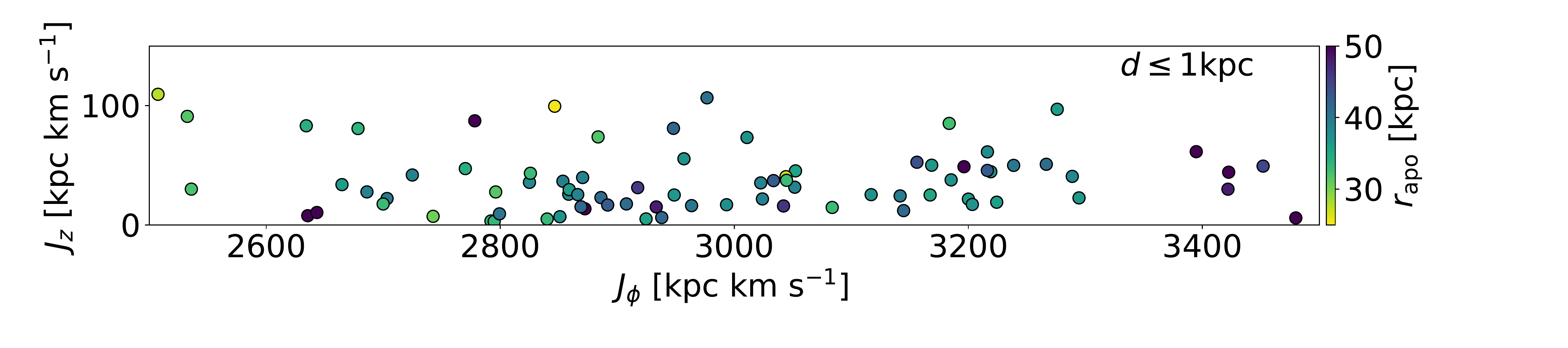}
   \caption{Selection procedure. Top panel: \textit{Gaia} DR3 stars from the selection process described in Section~\ref{sec:selection} (i.e. $\varpi/\delta \varpi>10$, $r_{apo}\ge 25\kpc$ and $d\le 1.5\kpc$). Middle panel: zoom on the low $J_z$ region delimited by the rectangle in the top panel ($2500\le J_\phi \le 3500 \kms \kpc$, $J_z\le150 \kms \kpc$). Bottom panel: same region as the middle panel, but for our final cut using distances $d\leq1\kpc$ from the Sun.}
\label{fig:SELECTION}
\end{figure*}

A similar technique was employed by \citet{2018MNRAS.478.5449M} to find several retrograde structures in the stellar halo, which were then tentatively associated to the $\omega$ Centauri globular cluster, which \citet{2012ApJ...747L..37M} had already suspected of bringing in such material. Retrograde structures have been linked to accretion events for a long time \citep{2007Natur.450.1020C}, and it has been confirmed by \citet{2017A&A...598A..58H} that the less bound stars in the halo are typically on retrograde orbits. \citet{2021MNRAS.500.3750S} also highlight the importance of the metal poor retrograde halo population for tracing the early building blocks of the galaxy.

\citet{2019MNRAS.488.1235M} reexamined the structures from \citet{2018MNRAS.478.5449M} and linked them to a substantial merger event they named Sequoia. The Sequoia progenitor galaxy could have brought those retrograde groups and possibly $\omega$ Centauri as well. The fact that its stellar population is distinct in metallicity and orbital parameters from the Gaia-Sausage makes the event another important piece of the stellar halo puzzle.

\begin{figure*}[!htb]
\centering
  \includegraphics[angle=0,  clip, width=\textwidth]{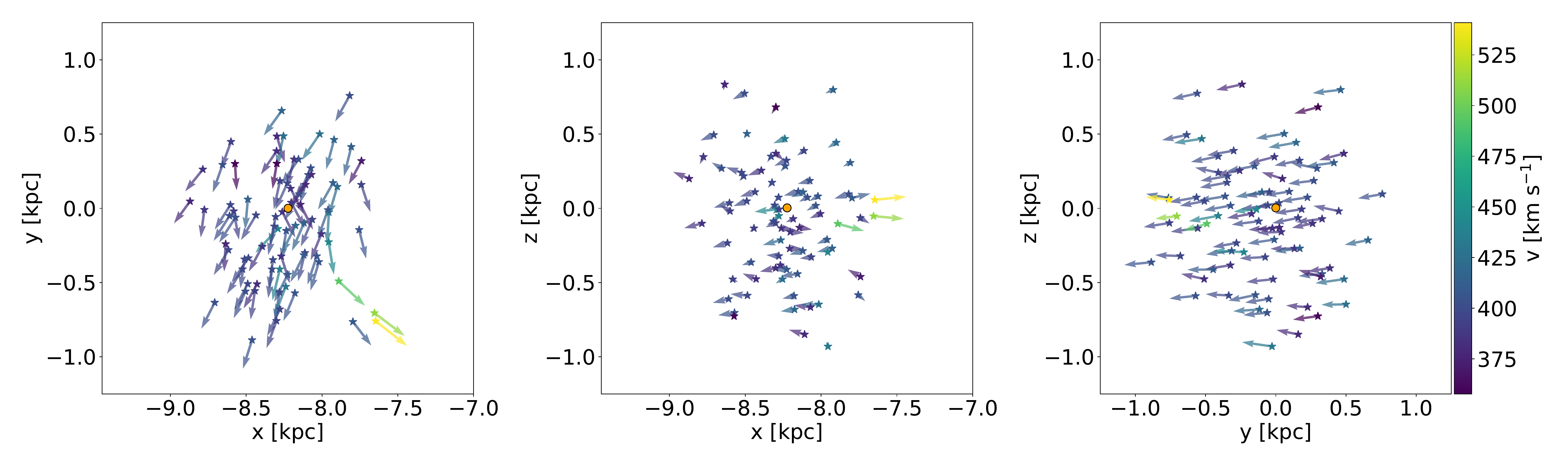}
    \includegraphics[angle=0, viewport= 30 0 1810 580, clip, width=\textwidth]{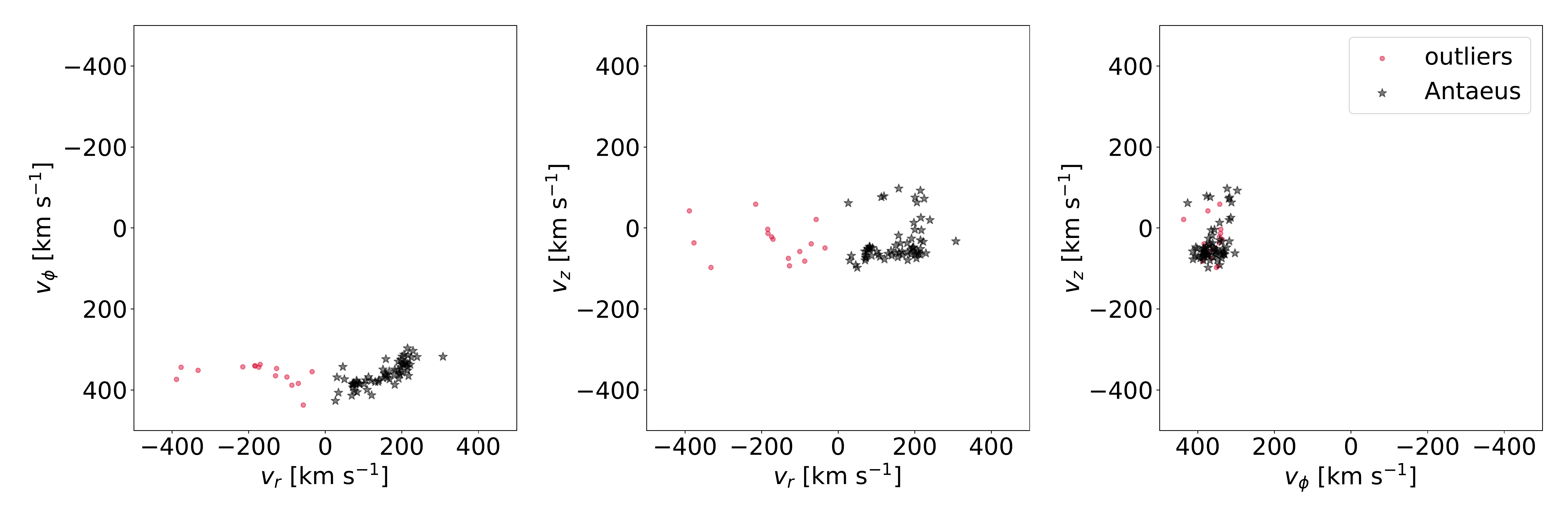}
   \caption{Top panel: position and velocity vectors of our selection of stars from Section~\ref{sec:selection} colored by total velocity; we plot bulk motion outliers with a slightly transparent line. The orange ball represents the Sun. Antaeus stars are currently passing through our Solar neighbourhood, going in a retrograde motion in the Milky Way's disk plane. Bottom panel: velocity planes $v_r v_\phi$, $v_r v_z$, $v_\phi v_z$ with the outliers (red dots) from the top panel bulk motion separated from Antaeus' stars (black). Note that we inverted the $v\phi$ axes to be coherent with usual velocity plots.}
\label{fig:BULK_MOTION}
\end{figure*}

In this work we present the discovery of Antaeus\footnote{In Greek mythology, Antaeus is the child of Gaia and Poseidon, a giant whose name comes from ``opponent''.}, a retrograde high energy group of tidal debris in the MW's disk plane, made using action-angle coordinates derived from the \textit{Gaia} DR3 catalogue \citep{GaiaDR3} and the St\"ackel fudge implemented in \textsc{Agama} \citep{2019MNRAS.482.1525V}. The new structure has several properties which are similar to those of Sequoia stars, so we discuss its possible affiliation to this event, although both its position in the disk of the MW and its extraordinary low vertical action make it stand out.

\section{Selection process}
\label{sec:selection}

Throughout this article, we use the right-hand side Galactic Cartesian coordinates for the MW with the Sun located at $(x,y,z)_\odot=(-8.2240,0,0.0028)\kpc$ (taking the Solar radius from \citealt{2020arXiv201202169B} and the height above the mid-plane from \citealt{2021A&A...646A..67W}) having peculiar velocity $(v_x,v_y,v_z)_\odot=(11.10,7.20,7.25)\kms$ (\citealt{2010MNRAS.403.1829S}, but with the velocity in the direction of Galactic rotation taken from \citealt{2020arXiv201202169B}). Our starting point is the Radial Velocity Spectrometer (RVS) sample of \textit{Gaia} DR3, for which we derive action-angle coordinates $(J_r,J_\phi,J_z)$ and orbital parameters using \textsc{Agama} \citep{2019MNRAS.482.1525V} in the  MW gravitational potential of  \citet{2017MNRAS.465...76M}. From this catalogue, we take the stars with good parallax measurements ($\varpi/\delta \varpi \geq10$) and $d \leq 1.5\kpc$ so as to retain a good quality Solar neighborhood sample. Since our aim is to investigate the structures that are falling down onto the Milky Way, we choose to select stars with large apocenter distances, $r_{\rm apo} \geq 25\kpc$. These cuts leave us with 3624 stars; we plot the resulting selection in the $J_\phi J_z$ plane, coloured by $r_{\rm apo}$, in Figure~\ref{fig:SELECTION} (top panel).

Among the many interesting structures that stand out from this view, we focus our attention on the low $J_z$, retrograde moving group of stars delimited by the black rectangle ($2500\le J_\phi \le 3500 \kms \kpc$, $J_z\le150 \kms \kpc$), into which we zoom in Figure~\ref{fig:SELECTION} (middle panel). We notice a good agreement in apocenters for stars in this region, further suggesting the presence of a stellar structure with coherent motion. 

\begin{figure}[!htb]
\centering
  \includegraphics[angle=0,  clip, width=\hsize]{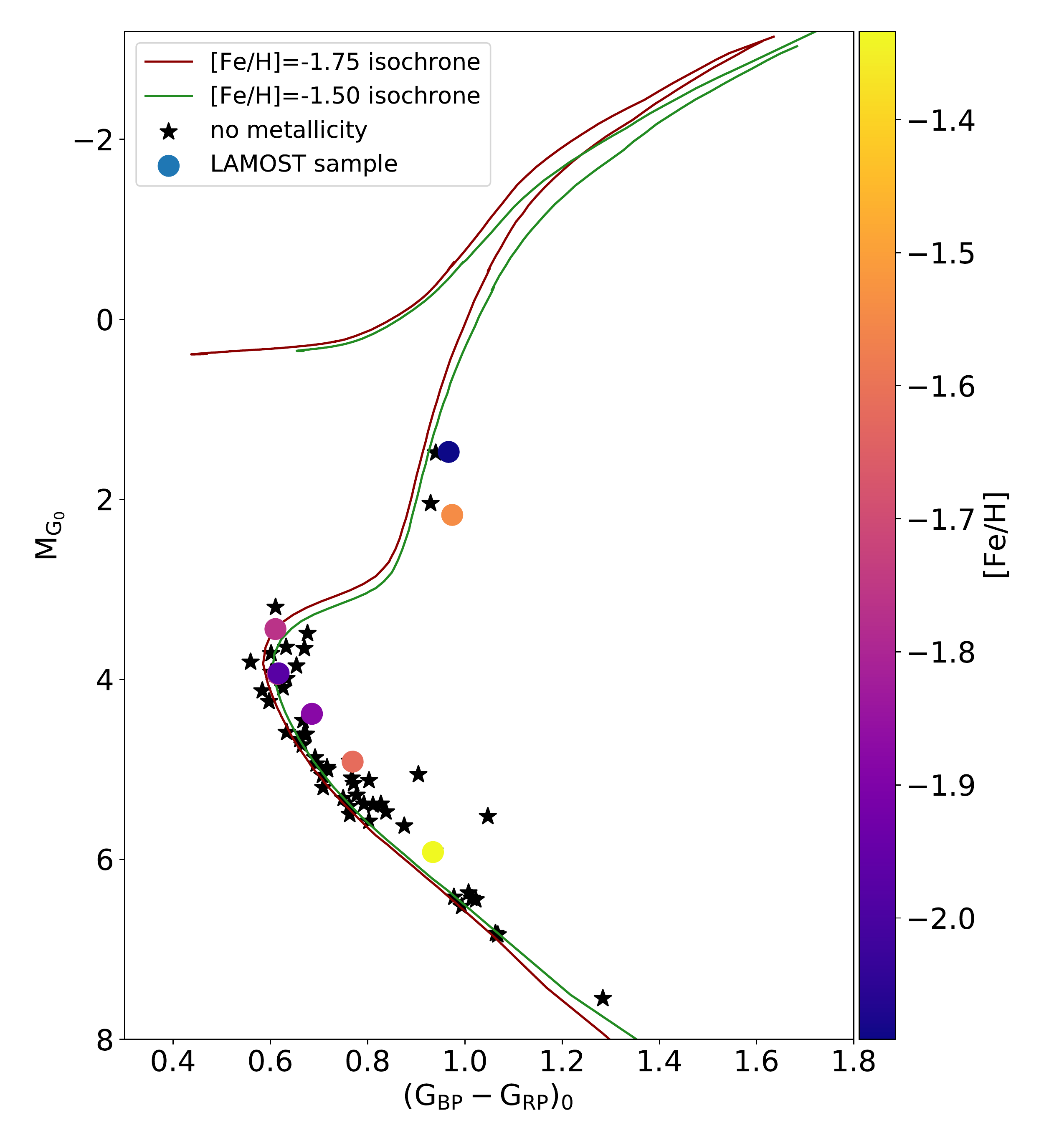}
   \caption{Colour magnitude diagram for our sample of Antaeus stars, compared to PARSEC model isochrones \citep{2012MNRAS.427..127B} of age $12\Gyr$ and metallicities ${\rm [Fe/H]=-1.75}$ (red) and ${\rm [Fe/H]=-1.50}$ (green). The colorbar gives the [Fe/H] for the 8 LAMOST stars.}
\label{fig:CMD}
\end{figure}

Finally, we experimented with the heliocentric distance cut to see how the selection changes. We noticed that by selecting stars within a distance of $d \leq 1\kpc$ from the Sun (Figure~\ref{fig:SELECTION}, bottom panel) the agreement in apocenters is slightly better, removing in particular some extreme values from the previous cut. This leaves a sample of 80 stars which are listed in Table~\ref{tab:data}.

\begin{figure*}[!htb]
\centering
  \includegraphics[angle=0,  clip, width=\textwidth]{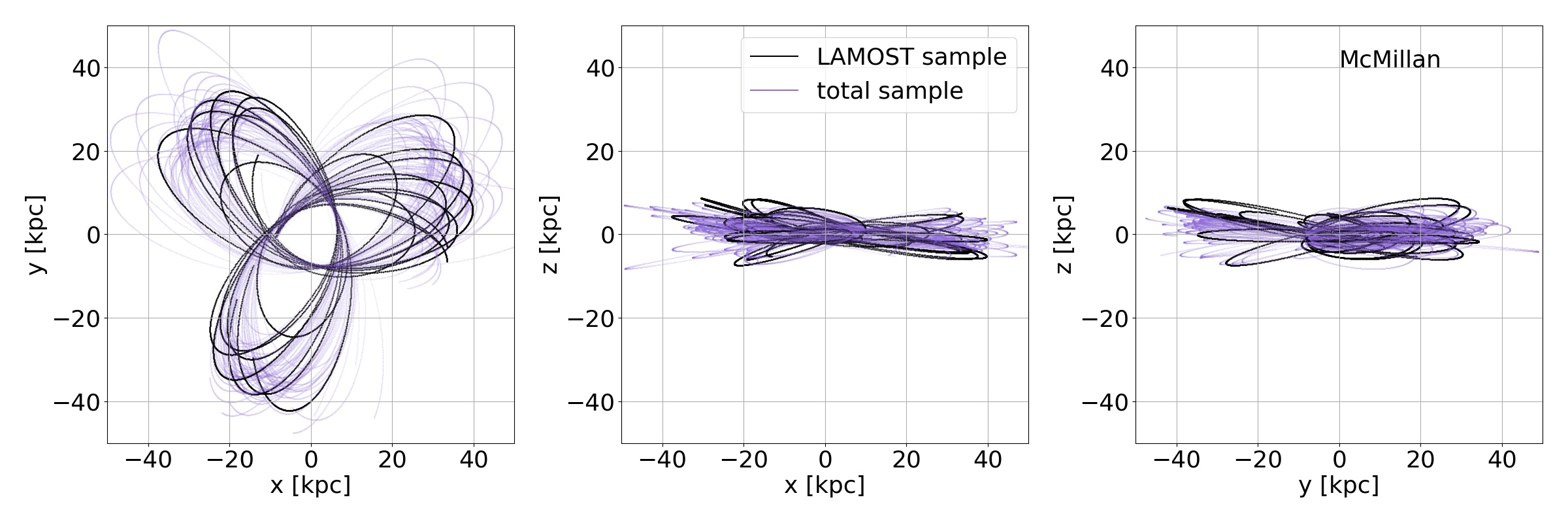}
  \includegraphics[angle=0,  clip, width=\textwidth]{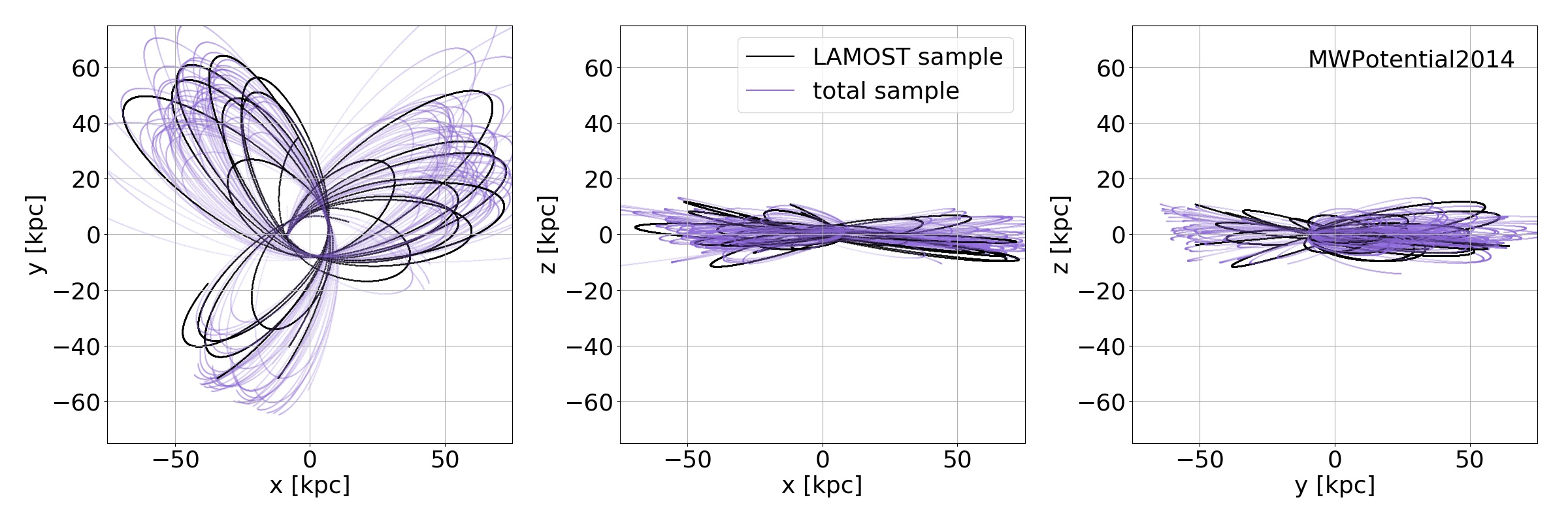}
   \caption{Orbits of Antaeus stars seen in Galactic Cartesian coordinates, integrated backwards in the \citet{2017MNRAS.465...76M} MW potential for $1.5 \Gyr$ (top panel), and in the MWPotential2014 model for $2.5 \Gyr$ (bottom panel). Notice the change of scales, as stars go farther when integrated in the lighter MWPotential2014. Orbits of the LAMOST sample (8 stars) are in solid black, and orbits of the rest of our sample (57 stars) are in purple.}
\label{fig:ORBIT}
\end{figure*}

\section{Sample characteristics}
\label{sec:characteristics}

We show the positions and velocities of our selection of stars in Figure~\ref{fig:BULK_MOTION} (top panel). It appears clear that the stars belong to a coherent structure dynamically, moving in a retrograde motion in the disk plane of the MW. The structure is rather thick, with a width of at least $1.5\kpc$. We identify some outliers from this bulk motion, which all have a distinctive positive velocity in the $x$ direction ($v_x\geq 0$). For the remainder of this study, we will exclude those 15 outliers from our sample, leaving us with 65 stars of the Antaeus stream. In Figure~\ref{fig:BULK_MOTION} (top panel), we plot velocity planes $v_rv_\phi$, $v_rv_z$, $v\phi,v_z$ with this separation taken into account, showing the compactness of Antaeus stars in those projections. 

We crossmatch our selection with the LAMOST DR8 catalogue \citep{2022ApJS..259...51W} and find 8 stars in common, for which we obtain metallicities from their ``FEH\_PASTEL'' values. These LAMOST stars have a mean ${\rm [Fe/H]}=-1.74^{+0.06}_{-0.07}$, with an intrinsic spread of $\sigma=0.11^{+0.10}_{-0.04}$ (correcting for the LAMOST metallicity uncertainty estimates) and individual values ranging from ${\rm [Fe/H]}=-1.33\pm0.23$ to ${\rm [Fe/H]}=-2.09\pm0.30$. 
The colour magnitude diagram (CMD) of the sample is shown on Figure~\ref{fig:CMD}, compared to old metal poor isochrones $(12\Gyr,\ {\rm [Fe/H]=-1.75} \text{ and } {\rm [Fe/H]=-1.50})$ from the PARSEC library \citep{2012MNRAS.427..127B}. The photometry is corrected for interstellar extinction using the 3D extinction estimates calculated by \citet{2022A&A...658A..91A}.

Finally, we integrate back in time the orbits of the Antaeus stars in the McMillan MW potential for $1.5 \Gyr$, and in the MWPotential2014 \citep{2015ApJS..216...29B}; we show the results in Figure~\ref{fig:ORBIT}. Here also the structure appears very coherent dynamically. We find, for the McMillan MW potential $(M_{\rm vir}=1.3\times10^{12}\msun)$, a mean pericenter radius of $r_{\rm peri} = 7.3\kpc$, a mean apocenter radius of $r_{\rm apo} = 39.3\kpc$, a mean orbital eccentricity of $ecc = 0.69$, and a mean orbital time of $t_{\rm orbit}=1.1\Gyr$. For the lighter MWPotential2014 however $(M_{\rm vir}=8\times10^{11}\msun)$, those values become mean $r_{\rm peri} = 7.3\kpc$, mean $r_{\rm apo} = 71.9\kpc$, mean $ecc = 0.81$, and mean $t_{\rm orbit}=1.5\Gyr$. The 8 LAMOST stars, whose orbits are plotted in solid black, appear to be good representative members of the stream.

The mean actions of stars in the structure are $(J_r=1761,J_\phi=2990,J_z=39)\kpc\kms$, and their mean energy is $E=-10^5$ km$^2$ s$^{-2}$ (in the \citealt{2017MNRAS.465...76M} potential model). 

\section{Discussion and Conclusions}
\label{sec:discussion}

Based on the characteristics derived in Section~\ref{sec:characteristics}, in particular the thickness of the structure (width $\simeq 1.5\kpc$) and the range of metallicity of its constituent stars, it seems highly likely that this group of stars is the remnant of a tidal stream of a disrupted dwarf galaxy. The CMD (Figure~\ref{fig:CMD}) seems to indicate that the progenitor is likely to be very old, probably around $\sim12\Gyr$ in age. The agreement is  better with a model metallicity of $\rm{[Fe/H]}=-1.50$, although we derive a mean value of ${\rm [Fe/H]}=-1.74^{+0.06}_{-0.07}$. It would thus be very helpful to extend our sample of metallicities to help decide the matter. 

The mean $J_\phi$, energy, and eccentricities of our sample of Antaeus stars show many similarities with the Arjuna/I'itoi/Sequoia group of mergers \citep{2020ApJ...901...48N}. However Antaeus seems more akin to the retrograde structures of \citet{2018MNRAS.478.5449M} and to the retrograde tail of the Sequoia event \citep{2019MNRAS.488.1235M}, especially when factoring in the metallicity of its population. The $\sim12\Gyr$ age derived from the CMD comparison is also consistent with estimates for Sequoia groups \citep{2022arXiv220102405R}. Nonetheless, Antaeus' extraordinarily low mean $J_z$ and its position in the disk plane of the MW both make it unique, even when compared to the global atlas of halo structures from \citet{2022ApJ...926..107M}. If the structure is indeed related to Sequoia, this difference has to be explained. 

The mere existence of such a streamy, retrograde structure in the disk of the MW is very puzzling. It is not clear how such kinematic coherence could be retained if this population came in with Sequoia $9\sim11\Gyr$ ago \citep{2019MNRAS.488.1235M}. Of course Antaeus' progenitor could have arrived initially with a small inclination, although this possibility appears somewhat contrived. It seems more natural to explain the very low quantity of vertical motion by dissipation due to dynamical friction, which might be consistent with an early arrival in the MW. This scenario would invite the possibility that Antaeus is the debris of the dense core of the Sequoia progenitor, which would have stabilized in the disk through dynamical friction before tidal disruption completely destroyed it. 

The discovery of Antaeus opens many exciting possibilities for follow-up studies. A first step would be finding other members of the structure in \textit{Gaia} with the information we now possess. Creating an $N$-body model for the infall of the progenitor dwarf galaxy in the potential well of the MW and exploring the possibilities for its survival in the disk would also be highly informative. Finally, it would be very helpful to measure the metallicity of more stars of our selection in order to facilitate discussions regarding the origin of the structure, and links to Sequoia in particular.

\section*{Data Availability}
\label{sec:Data}
The data used in this contribution is available in Table~\ref{tab:data}.

\begin{deluxetable*}{lcccccccccccc}
\centering
\tablehead{
    \colhead{\textit{Gaia} source ID} & \colhead{RA} &
    \colhead{DEC} & \colhead{$J_r$} & \colhead{$J_\phi$} & \colhead{$J_z$} & \colhead{$r_{\rm peri}$} & \colhead{$r_{\rm apo}$} & \colhead{$e$} & \colhead{[Fe/H]}\\
    \colhead{} & \colhead{[deg]} &
    \colhead{[deg]} & \colhead{$[\kpc\kms]$} & \colhead{$[\kpc\kms]$} & \colhead{$[\kpc\kms]$} & \colhead{$[\kpc]$} & \colhead{$[\kpc]$} & \colhead{} & \colhead{[dex]}
    }
    \tablewidth{\textwidth}
    \tablecaption{Sample from the 80 stars of our selection from Section~\ref{sec:selection}. The full sample is available in electronic format. \label{tab:data}}
\startdata
3857833427353671808&159.91&4.08&1843.09&106.55&2976.68&7.31&40.78&0.70&$-2.09\pm0.30$\\
1558668134509319040&204.99&49.77&698.65&99.50&2846.51&8.19&25.57&0.51&$-1.97\pm0.12$\\
1374889335770878848&232.59&35.38&1936.15&45.67&3216.25&7.72&43.03&0.70&$-1.88\pm0.10$\\
950636967397629568&102.78&40.33&1505.39&83.11&2634.22&6.55&34.23&0.68&$-1.77\pm0.13$\\
3839165510915273856&139.71&-0.89&1963.65&80.95&2947.99&7.09&41.87&0.71&$-1.76\pm0.08$\\
2657496656325125888&347.59&1.19&1963.74&9.28&2799.33&6.41&40.27&0.73&$-1.62\pm0.12$\\
231238462236707584&57.75&42.07&1680.13&33.76&2664.73&6.34&36.15&0.70&$-1.54\pm0.08$\\
3834229356541509760&147.87&0.89&1466.15&47.20&2770.15&6.84&34.32&0.67&$-1.33\pm0.23$\\
1950571427690143616&322.99&34.92&1216.32&14.63&3083.68&7.90&32.92&0.61&-\\
2340952515729081728&359.75&-21.81&1340.28&91.00&2532.56&6.35&31.65&0.67&-\\
\enddata
\tablecomments{The information is derived from the MW potential of \citet{2017MNRAS.465...76M}.}
\end{deluxetable*}

\begin{acknowledgments}
The authors acknowledge funding from the European Research Council (ERC) under the European Unions Horizon 2020 research and innovation programme (grant agreement No. 834148) and from the Agence Nationale de la Recherche (ANR projects ANR-18-CE31-0006 and ANR-19-CE31-0017). This work has made use of data from the European Space Agency (ESA) mission {\it Gaia} (\url{https://www.cosmos.esa.int/gaia}), processed by the {\it Gaia} Data Processing and Analysis Consortium (DPAC, \url{https://www.cosmos.esa.int/web/gaia/dpac/consortium}). Funding for the DPAC has been provided by national institutions, in particular the institutions participating in the {\it Gaia} Multilateral Agreement.

\end{acknowledgments}

\bibliography{disk_aliens}
\bibliographystyle{aasjournal}

\end{document}